\documentclass[preprintnumbers,amsmath,amssymb,floatfix,10pt,prd,
onecolumn,showpacs,superscriptaddress,nofootinbib]{revtex4}
\usepackage{graphicx}
\usepackage{epsfig}
\usepackage{bm}
\usepackage{amsfonts}

\begin{document}

\title{Observational constraints on non-minimally coupled Galileon model  }

\author{\textbf{ Mubasher Jamil}} \email{jamil.camp@gmail.com}
\affiliation{Center for Advanced Mathematics and Physics (CAMP),\\
National University of Sciences and Technology (NUST), H-12,
Islamabad, Pakistan}
\author{\textbf{Davood Momeni}}
\email{d.momeni@yahoo.com }
 \affiliation{Eurasian International Center
for Theoretical Physics,  L.N. Gumilyov Eurasian National
University, Astana 010008, Kazakhstan}
\author{\textbf{ Ratbay Myrzakulov}}
\email{rmyrzakulov@gmail.com}\affiliation{Eurasian International
Center for Theoretical Physics,  L.N. Gumilyov Eurasian National
University, Astana 010008, Kazakhstan}

\begin{abstract}
{\bf Abstract:} As an extension of Dvali-Gabadadze-Porrati (DGP)
model, the Galileon theory has been proposed to explain the
``self-accelerating problem'' and ``ghost instability  problem''. In
this Paper, we extend the Galileon theory by considering a
non-minimally coupled Galileon scalar with gravity. The statefinder analysis, $Om(z)$ diagnostic and
constraint the model parameters have been investigate 
, we find
$\Omega_\text{m0}=0.263_{-0.031}^{+0.031}$, $n=1.53_{-0.37}^{+0.21}$
(at the $95\%$ confidence level) with $\chi^2_{\text{min}}=473.376$.
Further we show  that  due to the SNe Ia + BAO data ,our model
behaves like a phantom-like dark energy.
\end{abstract}
%\pacs{04.20.Fy; 04.50.+h; 98.80.-k}
 \maketitle

%%%%%%%%%%%%%%%%%%%%%%%
\section{Introduction}
%%%%%%%%%%%%%%%%%%%%%%%

The observational data can be use to probe the the equation of state (EoS) of dark
energy $w_{\rm X}$ using the Supernovae Ia (SNe Ia) data \cite{SNearly} , Cosmic
Microwave Background radiations (CMB) \cite{CMBearly} and Baryon
Acoustic Oscillations (BAO) \cite{BAO1,Percival:2009xn}.

Over the past decade,different kinds of the dynamical dark energy
models have been discussed  (see Refs.~\cite{review} for review).
 Some popular models are like quintessence \cite{quin},
$f(R)$gravity \cite{fR}, scalar field models \cite{stensor}, the
Dvali-Gabadadze-Porrati (DGP) braneworld \cite{DGP}
scenario,modified gravities \cite{Tsujikawa:2010zza}, the
Gauss-Bonnet gravity \cite{fG}, $f(R,{\cal G})$ gravity
\cite{fRG},$f(R,T)$ gravity (Here $T$ is the trace of the
energy-momentum tensor), \cite{frt,ft} and so on.
 Physically we need to
find an effective gravitational action, which can recover the
Einstein gravity \cite{Will}.
These modifications must be free from any extra degree of freedoms
due to the ghost\cite{DeFelice,DGPghost}. In modified $f(R)$ theory
and it's reduction to scalar field models, we need to be careful in
picking the mathematical forms of $f(R)$ or the field potentials
functions in order to have compatibility with astrophysical
observations \cite{fRviable}.
 The scalar mode of the DGP theory is due to a longitudinal mode of a free massless sping 2 graviton with self interaction $\square \phi (\partial^{\mu} \phi
\partial_{\mu} \phi)$ which is the
 mixing with the transverse graviton\cite{DGPnon}.
 The physical  mechanism which
 is hidden behind such decoupling  is so-called by
Vainshtein mechanism \cite{Vainshtein}. It means that it is possible
to recover the Einstein gravity in a region of spacetime in size of the solar scales. The graviton interaction term of the form $\square
\phi (\partial^{\mu} \phi \partial_{\mu} \phi)$ satisfies the
non-Lorentzian invariance form of the  classical boost symmetry,
resembles  the Galilean local boost transformation
$$
\nabla_{\mu}\phi\rightarrow \nabla_{\mu}\phi+c_\mu,
$$
in the flat space-time.
The non relativistic model, based on the Galilean symmetry called as
the ``Galileon'' \cite{Nesseris:2010pc}. It has been shown that
\textit{there are only five field Lagrangians} ${\cal L}_i$
($i=1,\cdots, 5$) which are invariant under  the Galilean symmetry.
Their discussion was based on the Minkowski background. The equation
of the motion (EOM) derived from this action is second-order.
Consequently, the model seems free from extra unphysical degenerated
modes.
The plan of this Paper is the following: In section II, we introduce
our proposed model of non-minimal Galileon cosmology. In section
III, we perform the statefinder and Om diagnostics on model. In
section IV, we discuss the observational constraints on our model.
In section V, we provide the Conclusion of our Paper.

%%%%%%%%%%%%%%%%%%%%%%%%%%%%
\section{Non-minimal Galileon Cosmology}
%%%%%%%%%%%%%%%%%%%%%%%%%%%%

The covariant Galileon action reads \cite{Nesseris:2010pc}
\begin{equation}
\mathcal{S}=\int {\rm d}^4 x \sqrt{-g}\,\left[
\frac{1}{2\kappa^2}(1-\epsilon\kappa^2\xi\pi^2)R+ \frac12
\sum_{i=1}^5 c_i {\cal L}_i \right] +\int {\rm d}^4 x\, {\cal
L}_{M}\,, \label{action}
\end{equation}
with $g$ as $det(g_{\mu \nu})$ in units of $\kappa^2=8\pi G$, and the Galileon coupling constants  $c_i$ are constants. The covariant
Lagrangians ${\cal L}_i$ ($i=1, \cdots, 5$)  are given
by \cite{Nesseris:2010pc}
\begin{eqnarray}
& & {\cal L}_1=M^3 \pi\,,\quad {\cal L}_2=(\nabla \pi)^2\,,\quad
{\cal L}_3=(\square \pi) (\nabla \pi)^2/M^3\,, \nonumber \\
& & {\cal L}_4=(\nabla \pi)^2 \left[2 (\square \pi)^2 -2 \pi_{;\mu
\nu} \pi^{;\mu \nu}-R(\nabla \pi)^2/2 \right]/M^6,
\nonumber \\
& & {\cal L}_5=(\nabla \pi)^2 [ (\square \pi)^3
-3(\square \pi)\,\pi_{; \mu \nu} \pi^{;\mu \nu} \nonumber \\
& &~~~~~~~+2{\pi_{;\mu}}^{\nu} {\pi_{;\nu}}^{\rho}
{\pi_{;\rho}}^{\mu} -6 \pi_{;\mu} \pi^{;\mu \nu}\pi^{;\rho}G_{\nu
\rho} ] /M^9\,, \label{lag}
\end{eqnarray}
where $M$  is the mass parameter of  Galileon model.
Using the following standard metric
\begin{equation}
{\rm d}s^2=-{\rm d}t^2+a^2(t) \left[ \frac{{\rm d} r^2}{1-Kr^2}+r^2
({\rm d}\theta^2+ \sin^2 \theta\,{\rm d}\phi^2) \right]\,,
\label{metric}
\end{equation}
the equations of motion read
\begin{eqnarray}
& & 3H^2=\kappa^2(\rho_{\rm DE}+\rho_m+\rho_r+\rho_K+\rho_\xi)\,,
\label{basic1} \\
& & 3 H^2+2 \dot{H}=-\kappa^2(P_{\rm DE}
+\rho_r/3+\rho_K/3+P_\xi)\,,
\label{basic2}\\
& & \dot{\rho}_m+3H\rho_m=0\,,
\label{basic3}\\
& & \dot{\rho}_r+4H\rho_r=0\,,
\label{basic4}
\end{eqnarray}
where
\begin{eqnarray}
&&\rho_\xi=6\xi\epsilon\pi H\dot H+3\xi\epsilon\pi^2 H^2,\\
P_\xi&=&\frac{\epsilon}{2}(1-4\xi)\dot\pi^2+2\epsilon\xi\dot\pi-2\epsilon\xi(1-6\xi)\dot
H\pi^2\\&& \nonumber-3\xi\epsilon(1-8\xi)H^2\pi^2-\frac{1}{2}c_1\xi\pi.
\end{eqnarray}
and $\rho_K \equiv -3K/a^2\kappa^2$.
\begin{eqnarray}
\rho_{\rm DE} &\equiv& -c_1 M^3 \pi/2-c_2 \dot{\pi}^2/2
+3c_3 H \dot{\pi}^3/M^3 \nonumber \\
& &-45 c_4 H^2 \dot{\pi}^4/(2M^6)
+21c_5 H^3 \dot{\pi}^5/M^9,\\
P_{\rm DE} &\equiv&  c_1 M^3 \pi/2-c_2 \dot{\pi}^2/2
-c_3 \dot{\pi}^2 \ddot{\pi}/M^3 \nonumber \\
& &+3c_4 \dot{\pi}^3 [8H\ddot{\pi} +(3H^2+2\dot{H})
\dot{\pi}]/(2 M^6) \nonumber \\
& & -3c_5 H \dot{\pi}^4 [5H \ddot{\pi}+2(H^2+\dot{H}) \dot{\pi}
]/M^9\,.
\end{eqnarray}
The solutions of (\ref{basic3}) and (\ref{basic4}) are respectively
given by
\begin{equation}
\rho_\text m=\rho_\text{m0}a^{-3},\ \ \rho_\text
r=\rho_\text{r0}a^{-4}.
\end{equation}
 We must write (\ref{basic2}) in the following form
\begin{equation}
(H^2)'=-\kappa^2\Big(\rho_{\rm DE}+P_{\rm DE}
+\frac{4}{3}\rho_m+\rho_\xi+P_\xi\Big),\label{sys}
\end{equation}
in this new representation the deprivates are written with respect
to the e-folding   $\mathcal{N}=\ln a$. Figure 1 shows the time
evolutionary scheme of the metric and the Galileon gauges,
numerically. Also, the agreement of the Hubble parameter in our
model and LCDM model is obviously manifested from the right panel.

\begin{figure*}[thbp]
\begin{tabular}{rl}
\includegraphics[width=7cm]{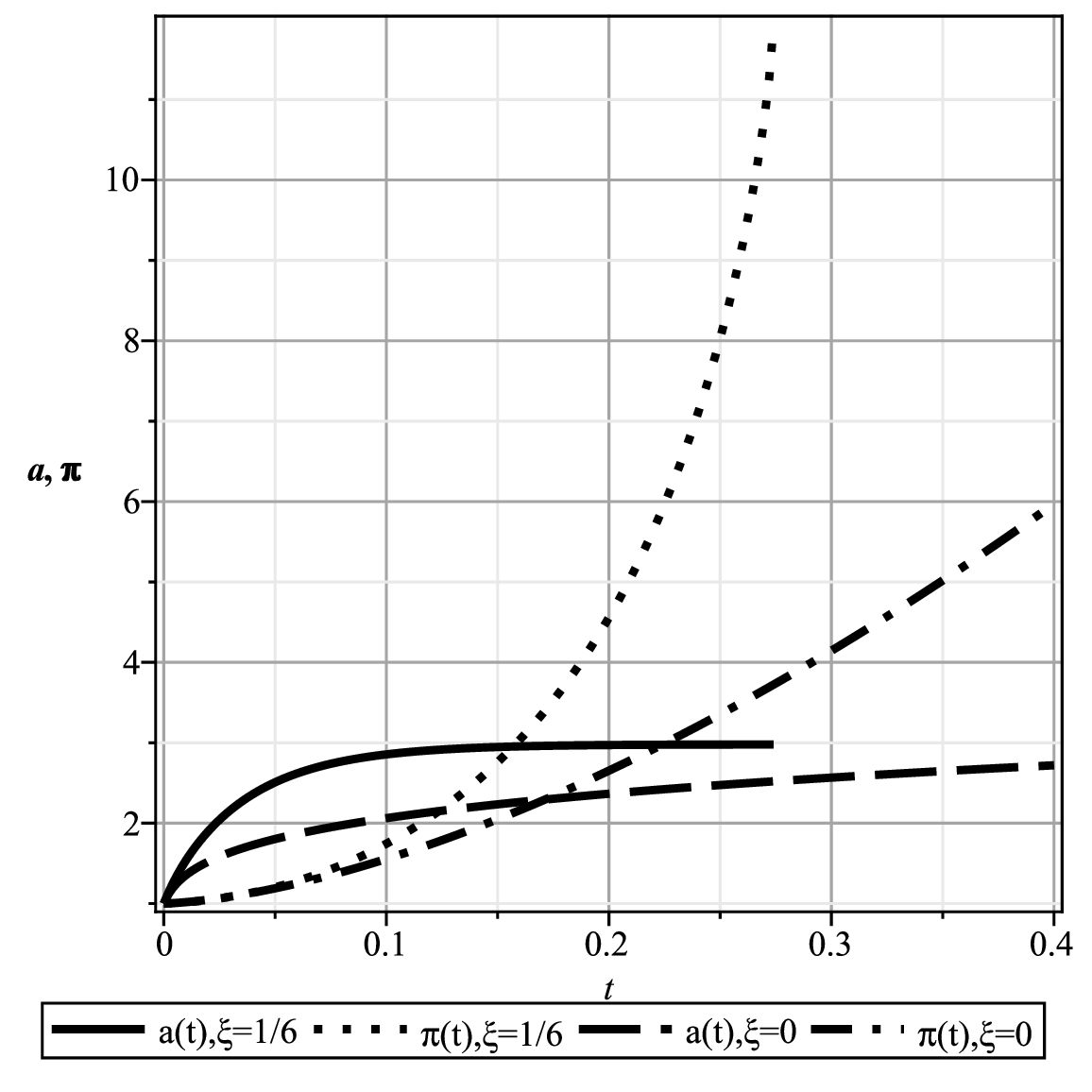}&
\includegraphics[width=7cm]{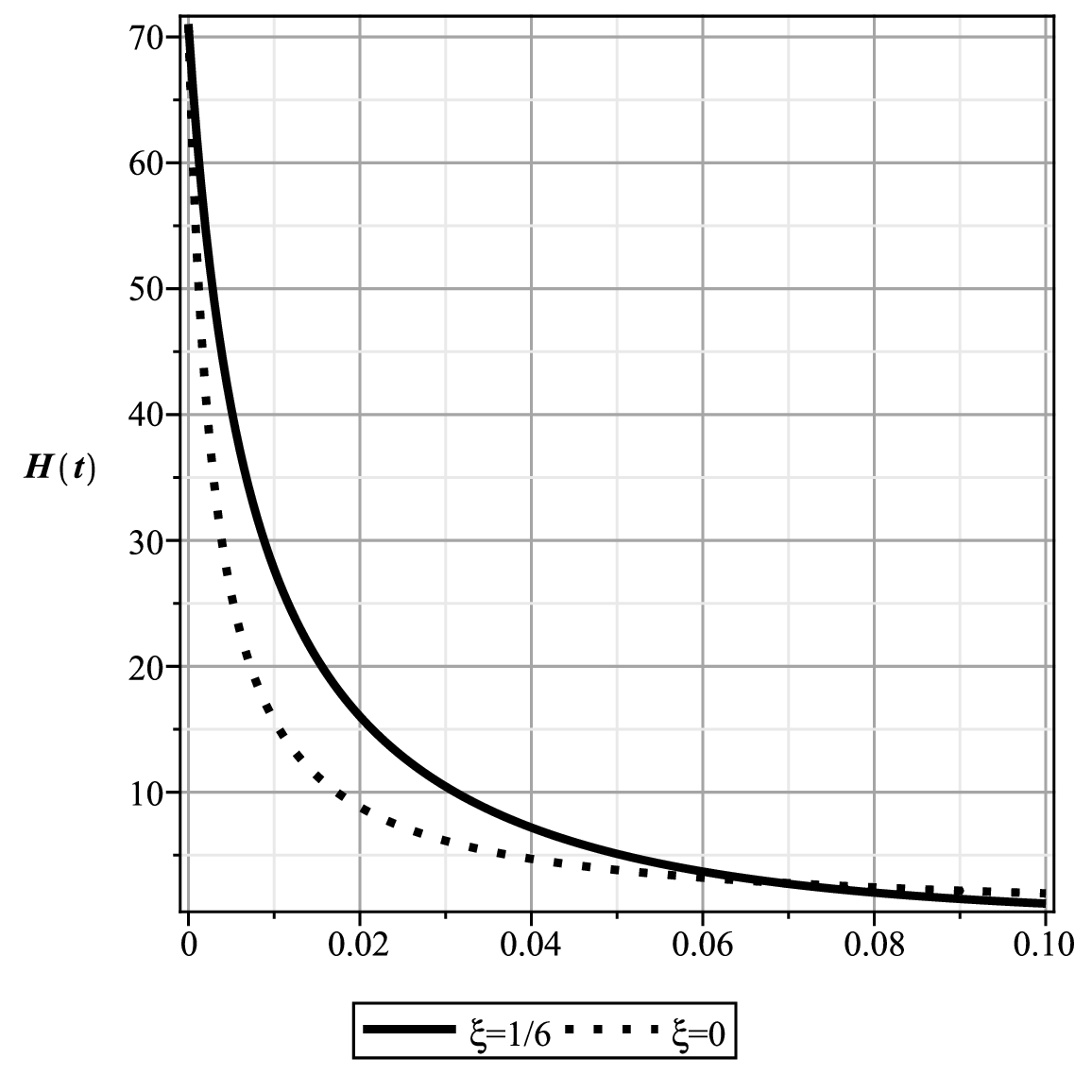} \\
\end{tabular}
\caption{(\textit{Left}) Time evolution of scale factor $a(t)$ and
Galileon scalar $\pi$. (\textit{Right}) Time evolution of the Hubble
parameter $H(t)$.}
\end{figure*}

\begin{figure*}[thbp]
\begin{tabular}{rl}
\includegraphics[width=7cm]{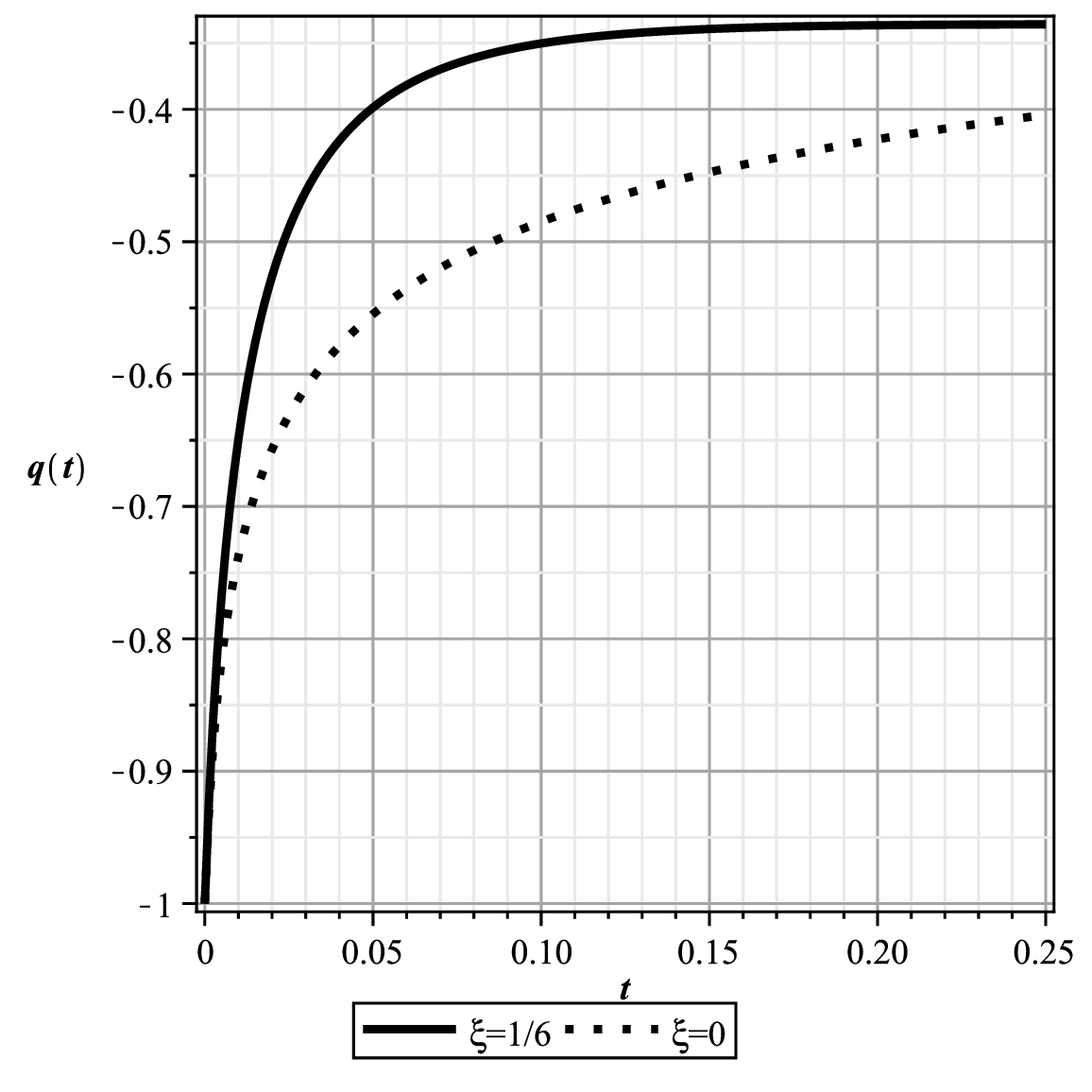}&
\includegraphics[width=7cm]{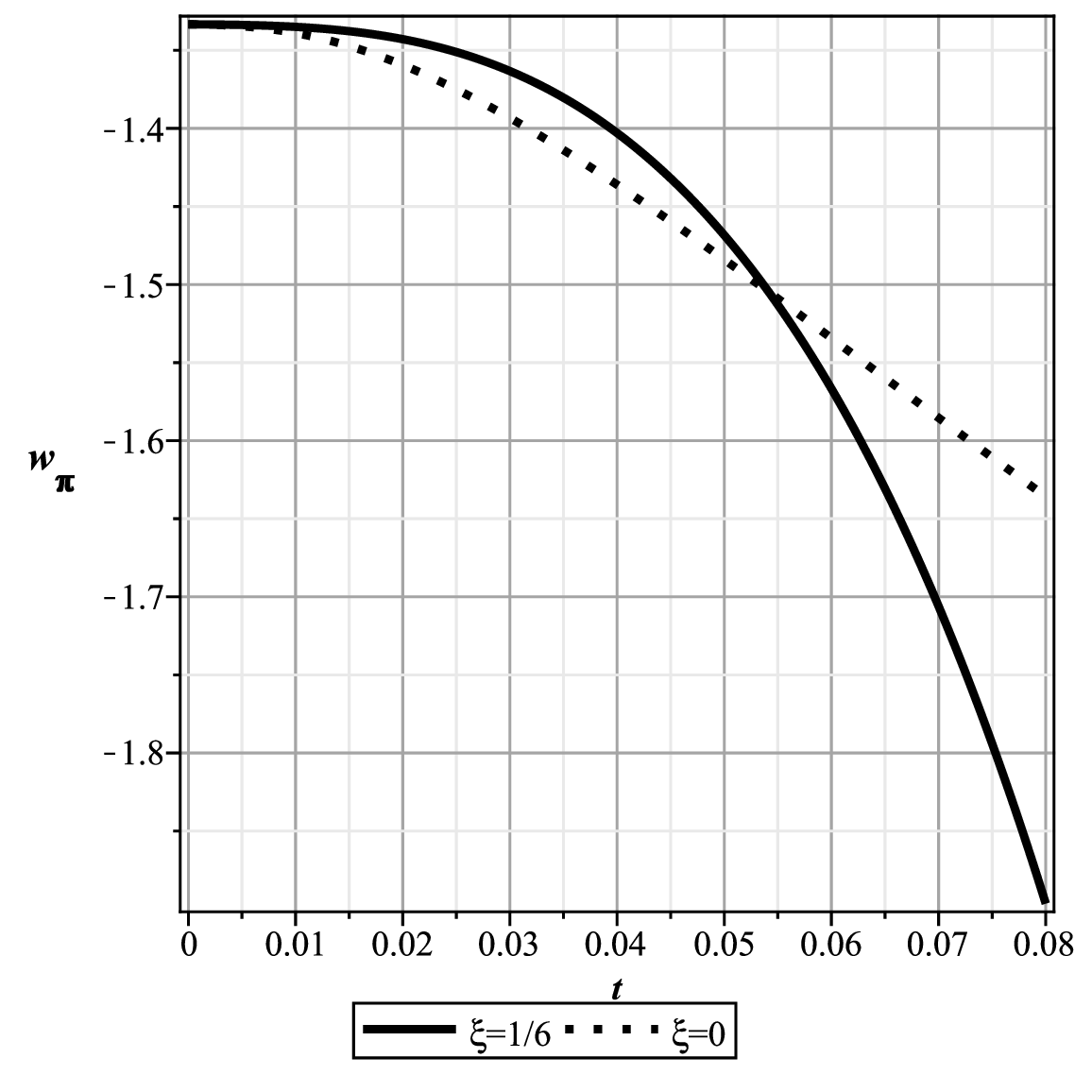} \\
\end{tabular}
\caption{(\textit{Left}) Deceleration parameter
$q$.  (\textit{Right}) Equation of state parameter
$w_\pi$. }
\end{figure*}
%%%%%%%%%%%%%%%%%%%%%%%%%%%%%%%%%%%%%%%%%%%%%%%%%%%%
\section{Statefinder Analysis and $Om$ Diagnostic}
%%%%%%%%%%%%%%%%%%%%%%%%%%%%%%%%%%%%%%%%%%%%%%%%%%%%%

In order to classify the different dark energy models, Sanhi
\textit{et al.} \cite{Sanhi} proposed a geometrical diagnostic
method by considering higher derivatives of the scale factor. The statefinder parameters $\{r, s\}$
 are  defined
\begin{eqnarray}
r\equiv \frac{\dddot{a}}{aH^3} \;,
\end{eqnarray}
\begin{eqnarray}
s\equiv \frac{r-1}{3(q-1/2)} \;,
\end{eqnarray}
where $q\equiv -\frac{1}{H^2}\frac{\ddot{a}}{a}$ is the deceleration
parameter. Apparently, $\Lambda$CDM model corresponds to a point
$\{1,0\}$ in $\{r, s\}$ phase space. The statefinder diagnostic can
discriminate different models. For example, it can distinguish
quintom from other dark energy models~\cite{WuYu}. From the panel of
figure-1, we observe that behavior of Hubble parameter can be
approximated as
\begin{equation}
H\propto\frac{1}{t^n},\ \ n\geq1.\label{hn}
\end{equation}
With this ansatz form, the behavior of statefinder parameters is
\begin{equation}
r=\frac{n^2-3n+2}{n^2},\ \ s=\frac{2(3n-2)t^n}{3n(nt^n-2)}.
\end{equation}
For very far future $t\rightarrow\infty$,
\begin{equation}
r=1-\frac{3n-2}{n^2},\ \ s=\frac{2(3n-2)}{3n^2},
\end{equation}
which can be combined as
\begin{equation}
r=1-\frac{3s}{2}.
\end{equation}
From (\ref{hn}), the scale factor  evolves like
\begin{equation}
a(t)=a_0\exp\Big( \frac{t^{1-n}-t_0^{1-n}}{1-n} \Big)\label{an}.
\end{equation}
From this expression we obtain the Hubble parameter $H(z)$:
\begin{equation}
H(z)=H_0\Big[1+t_0^{n-1}(n-1)\log(1+z)\Big]^{\frac{n}{n-1}}\label{hz}.
\end{equation}
We assume that $t_0=1$. So, the parameter for our model is $n$.
Indeed, $n=n(\xi)$. We interpolate
\begin{equation}
n(\xi)=\Sigma_{m=1}^{5}c_m \xi^{m},\ \ \xi\in \textit{Z}\label{nxi}.
\end{equation}
It is very interesting to investigate the behavior of the (\ref{an})
in limit of the limit $n\rightarrow1$. In this case for (\ref{an})
we have
\begin{eqnarray}
a(t)=a_0\Big(\frac{t}{t_0}\Big),\ \ H(t)=\frac{1}{t}.
\end{eqnarray}

The $Om(z)$ is another diagnostic of dark energy proposed by Sahni
\textit{et al.}~\cite{Sahni2008}.  It is defined as
\begin{equation}
Om(z)\equiv\frac{E^2(z)-1}{(1+z)^3-1}.
\end{equation}
By defining $E^2=H^2/H^2_0$. Obviously, this diagnostic parameter
depends only to the first derivative of the luminosity distance
$D_L(z)$.
  We are able from this diagnostic to discriminate
different dark energy  models by interpolating  the geometrical slope of $Om(z)$ although
we don't know the precise value of $\Omega_\text{m0}$. The figure 2 shows different graphs of the deceleration parameter and effective EoS  by indicating the DE behavior in the phantom era. Also the statefinder analysis has been presented in the figure 3.
\begin{figure}[htbp]
\includegraphics[width=5cm]{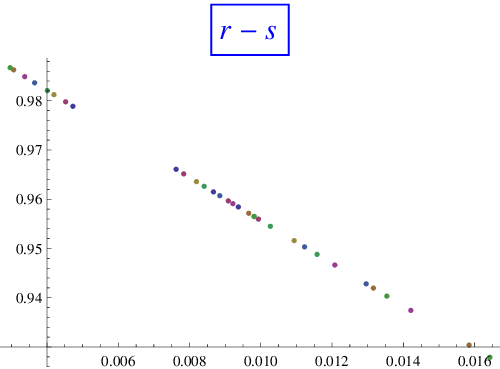}\includegraphics[width=5cm]{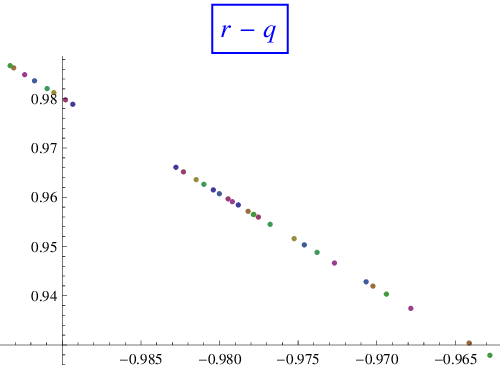}\includegraphics[width=5cm]{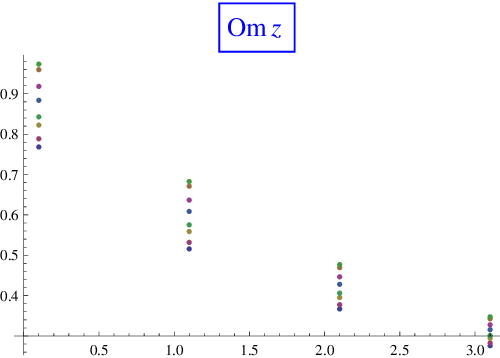}
\caption{\label{Figs1} The evolutionary curves of statefinder pair
$(r, s)$ (left), pair $(r, q)$ (middle) and $Om(z)$ (right) for
(\ref{hz}) with $\Omega_\text{m0}=0.278$, $-1<z<4$. Here we plot for
$1.2<n<1.5$ and $0.67<\xi<0.82$ }
\end{figure}

%%%%%%%%%%%%%%%%%%%%%%%%%%%%%%%%%%%%%%
\section{Observational Constraints}
%%%%%%%%%%%%%%%%%%%%%%%%%%%%%%%%%%%%%%5

We will now discuss the constraints on our model parameter $n$ which
appeared in (\ref{hz}) with (\ref{sys}). Here we perform the data analysis using SNe Ia, BAO and SDSS \cite{Amanullah2010}. First we must review these data sets (see Appendix A of \cite{Wu:2010mn} for a review).
\\
In (2010),  the Supernova  Cosmology Project collaboration
~\cite{Amanullah2010} reported the Union2 compilation, which
consists of 557 SNe Ia data points. In fact this is the largest
reported and spectroscopically confirmed SNe Ia sample . We use it
to constrain the theoretical models in this paper based on the model
(\ref{basic2}). As usually, the results  can be obtained by
minimizing the $\hat{\chi}^2$
\begin{eqnarray}
\hat{\chi}^2_{Sne}=\sum_{i=1}^{557}\frac{[\mu_{obs}(z_i)-\mu_{th}(z_i)]^2}{\sigma_{u,i}^2}\;,
\end{eqnarray}
where  $\sigma_{\mu,i}^2$ are the errors .
The luminosity distance $D_L$ can be calculated by
\cite{Eisenstein2005,Nesseris2005}
\begin{eqnarray}\label{dl}
D_L\equiv(1+z)\int_0^z\frac{dz'}{E(z')}.\;
\end{eqnarray}

Calculating the ${\chi}^2_{SNe}$, we find that,  the best fit values
are $\Omega_\text{m0}=0.271$, $n=n(\xi)=1.52$ with
$\chi^2_{Min}=481.272$. The results has been in figure 4 for
different confidence limits.

\begin{figure}[htbp]
\includegraphics[width=6cm]{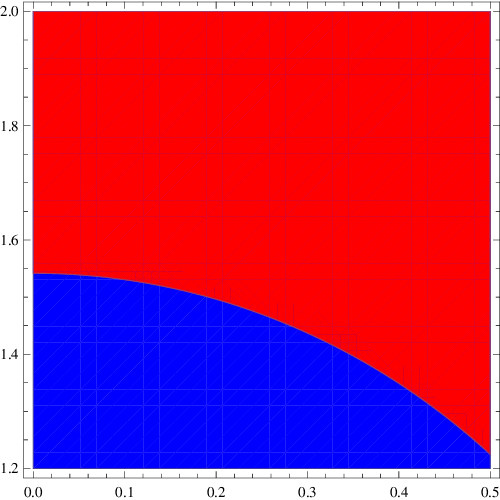}\includegraphics[width=6cm]{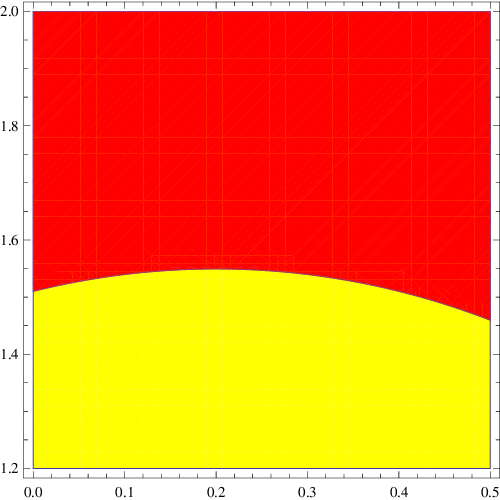}
 \caption{\label{Fig1} The $1\sigma$ and $2\sigma$ contours for $(\Omega_\text{m0},n)$ parameter space arising from the Sne Ia + BAO. The constraints on  Model (\ref{hz}) from  Sne Ia + BAO.
The  regions corresponds to $1-\sigma$ (Blue-left), $2-\sigma$
(Yellow-right)  confidence regions. Here Red is the background
color.}
\end{figure}
Now, using  BAO data. The parameter $A$ represented using the BAO
peak~\cite{Eisenstein2005}. The constraints from SNe Ia+BAO are
given by minimizing $\chi^2_{SNe}+\chi^2_{BAO}$. The results are
$\Omega_\text{m0}=0.263_{-0.031}^{+0.031}$, $n=1.53_{-0.37}^{+0.21}$
(at the $95\%$ confidence level) with $\chi^2_\text{min}=473.376$.
\newpage
%%%%%%%%%%%%%%%%%%%%%%%%%%%%%%%%
\section{Conclusions}
%%%%%%%%%%%%%%%%%%%%%%%%%%%%%%%%

We
constrained a  non-minimally coupled Galileon gravity with
Lagrangian
$\mathcal{L}=\frac{1}{2\kappa^2}(1-\epsilon\kappa^2\xi\pi^2)R+
\frac12 \sum_{i=1}^5 c_i {\cal L}_i$. Compared with references, we
examine our model with SNe Ia+BAO data . Using  SNe Ia and BAO, we find that the exponent power of Hubble parameter 
$n=\Sigma_{i=1}^5 c_m \xi^m\approx 1.5$, which contains the CDM
model.  We like to mention here that we have followed
largely the exposition given in the Wu and Yu paper \cite{wuyu1}.

%%%%%%%%%%%%%%%%%%

\end{document}